\documentclass[conference]{IEEEtran}
\usepackage{amsmath,bm}
\usepackage{graphicx}
\usepackage{graphics}
\usepackage{complexity}
\usepackage{multirow}
\usepackage{float} 
\usepackage{booktabs}
\usepackage[nolist]{acronym} 
\usepackage{hyperref}
\usepackage{comment}
\usepackage{amsfonts} 

\usepackage[acronym]{glossaries}

\newacronym{DUQ}{DUQ}{Deterministic Uncertainty Quantification}
\newacronym{CNN}{CNN}{Convolutional Neural Network}
\newacronym{CAD}{CAD}{Computer Aided Diagnosis}
\newacronym{OOD}{OOD}{Out of Distribution}
\newacronym{DNN}{DNN}{Deep Neural Network}
\newacronym{DNNs}{DNNs}{Deep Neural Networks}
\newacronym{IOD}{IOD}{In-Distribution}
\newacronym{SSDL}{SSDL}{Semi-supervised Deep Learning}
\newacronym{IID}{IID}{Independent and Identically Distributed}
\newacronym{FE}{FE}{Feature Extractor}
\newacronym{SSL}{SSL}{Semi-supervised Learning}

\newacronym{MCD}{MCD}{Monte Carlo Dropout}
\newacronym{MTCF}{MTCF}{Multi-Task Curriculum Framework}
\newacronym{CV}{CV}{Coefficient of Variation}
\newacronym{SDL}{SDL}{Supervised Deep Learning}

\ifCLASSINFOpdf
\else
\fi

\begin{document}

%
% paper title
% Titles are generally capitalized except for words such as a, an, and, as,
% at, but, by, for, in, nor, of, on, or, the, to and up, which are usually
% not capitalized unless they are the first or last word of the title.
% Linebreaks \\ can be used within to get better formatting as desired.
% Do not put math or special symbols in the title.
\title{Improving Semi-supervised Deep Learning by using Automatic Thresholding to Deal with Out of Distribution Data for COVID-19 Detection using Chest X-ray Images}

% Authors
\author{\IEEEauthorblockN{Isaac Benavides-Mata \IEEEauthorrefmark{2}\IEEEauthorrefmark{1}, Saul Calderon-Ramirez \IEEEauthorrefmark{2}\IEEEauthorrefmark{3} }

\IEEEauthorblockA{\IEEEauthorrefmark{2}Instituto Tecnologico de Costa Rica, Costa Rica}

\IEEEauthorrefmark{1}isaacbm@estudiantec.cr,
\IEEEauthorrefmark{3}sacalderon@itcr.ac.cr
}

% conference papers do not typically use \thanks and this command
% is locked out in conference mode. If really needed, such as for
% the acknowledgment of grants, issue a \IEEEoverridecommandlockouts
% after \documentclass

% for over three affiliations, or if they all won't fit within the width
% of the page, use this alternative format:
% 
%\author{\IEEEauthorblockN{Michael Shell\IEEEauthorrefmark{1},
%Homer Simpson\IEEEauthorrefmark{2},
%James Kirk\IEEEauthorrefmark{3}, 
%Montgomery Scott\IEEEauthorrefmark{3} and
%Eldon Tyrell\IEEEauthorrefmark{4}}
%\IEEEauthorblockA{\IEEEauthorrefmark{1}School of Electrical and Computer Engineering\\
%Georgia Institute of Technology,
%Atlanta, Georgia 30332--0250\\ Email: see http://www.michaelshell.org/contact.html}
%\IEEEauthorblockA{\IEEEauthorrefmark{2}Twentieth Century Fox, Springfield, USA\\
%Email: homer@thesimpsons.com}
%\IEEEauthorblockA{\IEEEauthorrefmark{3}Starfleet Academy, San Francisco, California 96678-2391\\
%Telephone: (800) 555--1212, Fax: (888) 555--1212}
%\IEEEauthorblockA{\IEEEauthorrefmark{4}Tyrell Inc., 123 Replicant Street, Los Angeles, California 90210--4321}}

% use for special paper notices
%\IEEEspecialpapernotice{(Invited Paper)}

% make the title area
\maketitle

% As a general rule, do not put math, special symbols or citations
% in the abstract
\begin{abstract}
Semi-supervised learning (SSL) leverages both labeled and unlabeled data for training models when the labeled data is limited and the unlabeled data is vast. Frequently, the unlabeled data is more widely available than the labeled data, hence this data is used to improve the level of generalization of a model when the labeled data is scarce. However, in real-world settings unlabeled data might depict a different distribution than the labeled dataset distribution. This is known as  distribution mismatch. Such problem generally occurs when the source of unlabeled data is different from the labeled data. For instance, in the medical imaging domain, when training a COVID-19 detector using chest X-ray images, different unlabeled datasets sampled from different hospitals might be used. In this work, we propose an automatic thresholding method to filter out-of-distribution data in the unlabeled dataset. We use the Mahalanobis distance between the labeled and unlabeled datasets using the feature space built by a pre-trained Image-net Feature Extractor (FE) to score each unlabeled observation. We test two simple automatic thresholding methods in the context of training a COVID-19 detector using chest X-ray images. The tested methods provide an automatic manner to define what unlabeled data to preserve when training a semi-supervised deep learning architecture.  
\end{abstract}

% no keywords

% For peer review papers, you can put extra information on the cover
% page as needed:
% \ifCLASSOPTIONpeerreview
% \begin{center} \bfseries EDICS Category: 3-BBND \end{center}
% \fi
%
% For peerreview papers, this IEEEtran command inserts a page break and
% creates the second title. It will be ignored for other modes.
\IEEEpeerreviewmaketitle

\section{Introduction}

\textit{\gls{DNNs}} have shown outstanding results on many supervised learning problems, but this practice requires large labeled datasets \cite{lecun2015deep}. This decreases considerably the number of problems that supervised learning can solve in many areas where labeled data is scarce. For example, the usage of labeled medical images requires expensive human effort (professional human annotators), thus building a labeled dataset is costly \cite{calderon2022dealing}.

To overcome this problem \textit{\gls{SSL}} algorithms have proven to be an alternative to building models less prone to over-fitting training data \cite{calderon2022_ssdl_mismatch}. This approach leverages both labeled and unlabeled data, providing a way to train models without the need of large labeled datasets \cite{zhu2005semi}. Unlabeled samples are in general easier to collect, as it usually does not require annotators. Hence, building an \textit{\gls{SSDL}} model can be a cheaper alternative to \textit{\gls{SDL}}.

Unlabeled data is available from many different sources. In semi-supervised learning, it is usually assumed that the labeled and unlabeled datasets follow a similar distribution. This is known as the \textit{\gls{IID}} assumption.  However, the risk that this assumption is violated under real-world usage settings can be high when using unlabeled data sampled from different sources. The phenomenon where the distribution of labeled dataset is not similar to the distribution in the unlabeled dataset is called \textit{distribution mismatch} \cite{oliver2018realistic}. An unlabeled observation that is very unlikely to belong to the labeled data distribution can be referred to as an \textit{\gls{OOD}} observation.

The accurate detection of COVID-19 is a critical task to control the pandemic. Fortunately, COVID-19 detection  using chest X-ray images can be considered an inexpensive method for its detection, as X-ray imaging systems are more widely available than other medical imaging technologies \cite{calderon2021improving}. Nevertheless, scarcely labeled data is a limitation faced to train a deep learning based COVID-19 detector \cite{Arora2014}. 

The problem of distribution mismatch between the labeled and unlabeled datasets for training a COVID-19 detector using chest X-ray was analyzed in \cite{calderon2022dealing}. In such work, a simple method to score each unlabeled chest X-ray image with respect to the labeled dataset was proposed. However, the method was tested using a previously known fixed threshold \cite{calderon2022dealing}. 

In this work, an automatic thresholding technique is proposed. This technique is used to filter the \gls{OOD} data in the unlabeled dataset, using the scores proposed in \cite{calderon2022dealing}.  The tested thresholding techniques use the scores to estimate the optimal threshold between the \textit{\gls{IOD}} and \gls{OOD} data. Furthermore, we propose a simple method to improve the data quality in the unlabeled dataset in a semi-supervised setting by removing the \gls{OOD} data. We evaluate the filter by using a small labeled dataset from the target clinic along with a large unlabeled dataset from the same or a different clinic. %estimate whether the unlabeled dataset depicts \gls{OOD} data or not.
%It is expected the improvement in the \gls{SSL} model’s accuracy as the harmful data in the unlabeled dataset has been removed.

\section{State of the Art}

In Section \ref{sec:SSL} we enlist recent major categories for \gls{SSL} for deep learning architectures. As the aim of this work refers to training \gls{SSL} methods under distribution mismatch settings, we later examine the literature around \gls{OOD} detection for deep learning models in Section \ref{sec:OOD}.

\subsection{Semi-supervised Learning approaches for Deep Learning Architectures in Image Analysis}\label{sec:SSL}

In \gls{SSL} the training datasets uses both labeled observations $S^{(l)}$ and unlabeled observations $S^{(u)}$. Using \gls{SSL} can be useful under real-world usage settings where collecting labeled data is expensive, or time-consuming.  \gls{SSL} leverages  unlabeled data which is frequently more widely available and less expensive to obtain \cite{ouali2020overview}. We refer to \gls{SSDL} when \gls{SSL} is implemented to train a set of layers in a deep learning model.

As developed in \cite{calderon2022_ssdl_mismatch,van2020survey} the most popular \gls{SSDL} architectures can be classified as pre-training based \cite{doersch2015unsupervised}, pseudo-labeled \cite{dong2018tri} and regularization-based \cite{berthelot2019mixmatch}.  In \cite{calderon2022_ssdl_mismatch} a detailed description of \gls{SSDL} methods can be found.
Pre-training based methods frequently implement a set of proxy tasks (e.g., the estimation of the image rotation angle) to pre-train one or more layers of the deep learning model \cite{simeoni2021rethinking}. Later the model is fine-tuned using the labeled data \cite{simeoni2021rethinking}. 

Pseudo-labeled methods estimate a set of  pseudo-labels, either by implementing an iterative process where the more confident estimations are re-used as labelled observations (e.g, boosting) or using a model ensemble \cite{van2020survey}. 

Regularization based \gls{SSDL}, also known as \textit{intrinsically unsupervised methods} \cite{van2020survey}, include an unlabeled data based term in the loss function $\mathcal{L}(S)$ to regularize the model during training. This is shown in Equation \ref{eq:loss}.

\begin{equation}
    \label{eq:loss}
    \mathcal{L}(S)=\sum_{(\bm{\mathit{x}}_i, \bm{\mathit{y}}_i)\in S^{(l)}} \mathcal{L}_l (\bm{\mathit{w}}, \bm{\mathit{x}}_i, \bm{\mathit{y}}_i) + \gamma \sum_{\vec{\mathit{x}}_j \in S^{(u)}} \mathcal{L}_u (\bm{\mathit{w}}, \bm{\mathit{x}}_j)
\end{equation}

where $\bm{\mathit{w}}$ is the model's weights array, and the labeled and unlabeled loss term are denoted by $\mathcal{L}_l$ and $\mathcal{L}_u$. The term $\gamma$ weights the influence of unsupervised regularization.

Regularization-based methodologies have been the most popular ones to implement \gls{SSL} for deep learning architectures \cite{calderon2022_ssdl_mismatch}. Earlier regularization-based approaches include the pseudo-ensemble \cite{bachman2014learning},  temporal ensembling \cite{laine2016temporal} and mean teacher \cite{tarvainen2017mean}. More recently, regularized methods like MixMatch combine regularized \gls{SSDL} with heavy data augmentation, yielding interesting results over previous architectures \cite{berthelot2019mixmatch}. MixMatch combines pseudo-labeling with regularization based \gls{SSL} by implementing the unlabeled loss term $\mathcal{L}_u$  with a comparison of the estimated label of the model with the average output of the model using a number of simple transformations (e.g, image flipping). This averaged model output is  referred to as a \textit{soft pseudo-label}.

MixMatch has been extended to include more sophisticated approaches of data augmentation in FixMatch \cite{sohn2020fixmatch} and ReMixMatch \cite{berthelot2019remixmatch}. Further within the MixMatch family, more recently in \cite{hu2021simple}, the MixMatch approach is modified to include a pair loss which minimizes the distance between observations with high confidence pseudo-labels with high similarity. The authors reported accuracy gains ranging from 1 to 3 percent with respect to FixMatch.

In this work we test our automatic \gls{OOD} scoring algorithm with the MixMatch method. However the proposed approach can be also combined with more recent developments within the MixMatch family.

\subsection{Out of Distribution Data Detection}\label{sec:OOD}
The task of detecting \gls{OOD} data is frequently approached as a score estimation problem. The score can be used as a measure to discern how likely  an observation belongs to  distribution of the labeled  dataset \cite{hendrycks2016baseline}.

In \cite{calderon2022dealing}, \gls{OOD} scoring methods are categorized as \textit{\gls{DNN}} output-based and feature or latent representation-based methods.

Output-based \gls{OOD} scoring methods use the \gls{DNN} output to estimate the \gls{OOD} likelihood. For instance \cite{hendrycks2016baseline} uses the softmax of the model's output as a \gls{OOD} score. Later, different calibration methods to improve the \gls{OOD} score mapping were developed. Also ensemble-based methods that use a number of model's perturbations as the \textit{\gls{MCD}} method developed in \cite{gal2016dropout} can be categorized as output-based. 

As for the feature space-based methods, Kimin et al. \cite{lee2018simple} proposed a simple method that compares the input observation to be scored with the training data, using the Mahalanobis distance.  Later, Joost et al. \cite{van2020simple} proposed a method referred as \textit{\gls{DUQ}}.  The proposed method computes the centroids for each class in the training dataset (\gls{IOD} dataset). For a new observation to be \gls{OOD} scored, the method calculates the distance to each centroid. The shortest distance is used as the uncertainty or \gls{OOD} score. 
% \todo{voy por aqui}
For instance, Qing et al. \cite{yu2020multi} proposed a \textit{\gls{MTCF}} that enables \gls{SSL} to achieve stable performance when the unlabeled dataset contains outliers by detecting them. The proposed method defines \gls{OOD} scores to the observations in the unlabeled dataset. The scores are optimized with the \gls{DNN} parameters, hence the scores are updated by using the \gls{DNN} output. To threshold the \gls{OOD} scores, the method uses the Otsu's thresholding method \cite{Otsu1979}. As a limitation, the thresholding method fails to improve the baseline accuracy when the number of \gls{IOD} and \gls{OOD} samples is imbalanced.

% \todo{OOD data scoring}

% \todo{THRESHOLDING}

\section{Proposed Method}
In this work, we propose an automatic thresholding technique to filter the \gls{OOD} data in the unlabeled dataset $S^{(u)}$. To do so, we estimate the harm coefficient of each unlabeled observation $\boldsymbol{s}^{(u)}$, using the method developed in \cite{calderon2022dealing}. The method uses the Mahalanobis distance, which assumes a Gaussian distribution in the feature space. The Mahalanobis distance is used as previous work suggests that the Mahalanobis distance is faster and slightly yields more accuracy gain than the feature histograms method. The feature space used is a generic feature space built from a pre-trained Image-net model using the AlexNet architecture, given its low computational cost. We summarize the proposed method in \cite{calderon2022dealing} as follows:

\begin{enumerate}
    \item Take the labeled observations in the matrix $S^{(l)}$, and the unlabeled observation ${\boldsymbol s}^{(u)}$ to score.  With a \textit{\gls{FE}} $h_{\mathrm{FE}}$ of an Image-net pre-trained Alexnet architecture, we calculate the set of feature arrays for the labeled observations $H^{(l)}=h_{\mathrm{FE}}\left(X^{(l)}\right)$, and the feature array for the unlabeled input observation ${\boldsymbol h}^{(u)}=h_{\mathrm{FE}}\left({\boldsymbol s}^{(u)}\right)$. The feature extractor we tested in this work yields $256$ features, thus ${\boldsymbol h} ^{(u)}\in\mathbb{R}^{256}$ and $H^{(u)}\in\mathbb{R}^{N\times256}$.
    
    \item As a next step, the Mahalanobis distance is calculated as follows:
    \begin{equation}
        \label{eqn:mahalanobis}
        d_{M}\left(\overline{\boldsymbol{h}}^{(l)},\boldsymbol{h}^{(u)}\right)=\left(\overline{\boldsymbol{h}}^{(l)}-\boldsymbol{h}^{(u)}\right)^{T}\Sigma_{l}^{-1}\left(\overline{\boldsymbol{h}}^{(l)}-\boldsymbol{h}^{(u)}\right)
    \end{equation}
    Where $\Sigma_{l}^{-1}$ corresponds to the covariance matrix of the labeled feature set $H^{(l)}$ and $\overline{\boldsymbol{h}}^{(l)}$ corresponds to the sample mean of such labeled feature set. 
    
\end{enumerate}

We define the histogram of the Mahalanobis distances of the observations within the unlabeled dataset as $\boldsymbol{p}^{(u)}_d$. The set of distances between each unlabeled observation $\boldsymbol{s}^{(u)}_j$ and the labeled dataset $S^{(l)}$ is defined by $D^{(u)}$.

The previous procedure shows the steps to calculate the set of distances $D^{(u)}$ to each observation in $S^{(u)}$. The larger distances are less favorable to use for training. Thus, when the unlabeled dataset has \gls{OOD} data, the histogram of distances $\boldsymbol{p}^{(u)}_d$ tends to form a bimodal Gaussian distribution, where one mode contains the \gls{IOD} data and the other mode the \gls{OOD} data. From there, we propose the usage of Otsu’s thresholding method and the K-means clustering, assuming one to two clusters, to find the optimal threshold in the distribution.

Otsu\textquotesingle s thresholding method is used to estimate an automatic thresholding in a histogram. In our case we use the previously estimated score density function $\boldsymbol{p}^{(u)}_d$. The method assumes a Gaussian distribution in the input histogram, hence the performance to estimate the optimal threshold depends that the input data follows the mentioned distribution function. The approach consist in estimating the threshold by maximizing the between-class variance $\sigma^2_b(\tau)$. A more detailed explanation can be found in \cite{Otsu1979}.

% \begin{equation}
% \label{eqn:interv}
% 	\sigma^2_b(\tau) = w_1(\tau) w_2(\tau) [\mu_1(\tau) - \mu_2(\tau)]^2
% \end{equation}

The K-means clustering estimates the nearest mean for the observations in the dataset. Therefore, the K-means algorithm is an iterative procedure that optimizes the means until stabilization is achieved \cite{gonzalez2008digital}. In our case we use as input for the K-means the scores dataset for the unlabeled observations $D^{(u)}$. This algorithm can be represented by the equation $
	\underset{C}{\arg \text{min}} (\sum_{i=0}^{k} \sum_{x \in C_i} || x - \mu_i ||^2)
$, where $x \in D^{(u)}$ is the set of scores, $k$ the number of clusters, and $C$ the clusters.

The aforementioned methods, estimate the threshold assuming the unlabeled dataset $S^{(u)}$ contains \gls{OOD} data. However, $S^{(l)}$ and $X^{(u)}$ may not depict distribution mismatch. As a result, the histogram $\boldsymbol{p}^{(u)}_d$ will present a mono-modal Gaussian distribution. Therefore, we propose to use the \gls{CV} to estimate when $\boldsymbol{p}^{(u)}_d$ depicts a mono-modal or a bi-modal distribution in order to select a dynamic threshold. The procedure is explained as follows:
%select whether the filter will preserve all the unlabeled datasets or it will filter the data using the selected thresholding method. This procedure is explained as follows:

\begin{enumerate}
    \item Compute the set of Mahalanobis distances $D^{(u)}$.
    
    \item Calculate the sample mean $\mu_\text{tot}$ and standard deviation $\sigma_\text{tot}$ from the set $D^{(u)}$.
    
    \item Obtain the threshold $\tau$ using the one of the thresholding methods.
    
    \item Using the threshold $\tau$ the set of distances $D^{(u)}$ is separated into two different sets:
    \begin{equation}
        D^{(u)}_{\text{lt}} = \{D^{(u)}_i \leq \tau, 1 \leq i \leq N\} \quad \text{and}
        % d^{u_1} = \text{find the elements in } d^{u} \text{ lower than } \tau
    \end{equation}
    \begin{equation}
        D^{(u)}_{\text{gt}} = \{D^{(u)}_i > \tau, 1 \leq i \leq N\} \qquad \
        % d^{u_2} = \text{find the elements in } d^{u} \text{ higher or equal than } \tau
    \end{equation}
    where N is the number of elements in $D^{(u)}$.
    
    \item Compute the sample mean and standard deviation from the sets, which are defined as $\mu_\text{lt}$ and $\sigma_\text{lt}$ for the set $D^{(u)}_{\text{lt}}$, and $\mu_\text{gt}$ and $\sigma_\text{gt}$ for the set $D^{(u)}_{\text{gt}}$.
    
    \item Calculate the \gls{CV}, as follows:
    \begin{equation}
        \text{CV}_\text{tot} =  \frac{\sigma_\text{tot}}{\mu_\text{tot}}, \qquad \text{CV}_{u_\text{lt}} =  \frac{\sigma_\text{lt}}{\mu_\text{lt}}, \qquad \text{CV}_{u_\text{gt}} =  \frac{\sigma_\text{gt}}{\mu_\text{gt}}
    \end{equation}
    % \begin{equation}
    %     \text{CV}_{u_1} =  \frac{\sigma_1}{\mu_1}
    % \end{equation}
    % \begin{equation}
    %     \text{CV}_{u_2} =  \frac{\sigma_2}{\mu_2}
    % \end{equation}
    
    \item Compare the \gls{CV} of $\text{CV}_\text{tot}$, and $\text{CV}_{u_\text{lt}}$ and $\text{CV}_{u_\text{gt}}$:
    \begin{equation}
        \label{eq:cv}
        \alpha \cdot \text{CV}_\text{tot} < \text{CV}_{u_\text{lt}} + \text{CV}_{u_\text{gt}}
    \end{equation}
    
    \item In case the previous step is false then the filter assumes that all the data in the unlabeled dataset is \gls{IOD} and the filter's output will correspond to the observations with distance $D^{(u)}$. Otherwise, the filter's output are the observations which distance is contained in $D^{(u)}_{\text{lt}}$.
\end{enumerate}

%assuming one or two mode 1 distribution or cluster and the assumption of 2 distributions or clusters.
In step 6, the \gls{CV} is a statistical measure that estimates the dispersion of data in a cluster. This coefficient is used to compare the variability of data between $D^{(u)}$, and the sets $D^{(u)}_{\text{lt}}$ and $D^{(u)}_{\text{gt}}$. For this reason, in step 7 both coefficients of variation are compared to choose the best-suited assumption. Moreover, in step 7 a constant $\alpha$ is defined, this value is used to increase a little the value of $\text{CV}_\text{tot}$. The suggested values of $\alpha$ should be around around $1.12$. This value was chosen due experimentation.

%$1 < \alpha < 1.3$, for instance the value used for the experiments was

%filter of \gls{OOD} data can improve the accuracy of a model trained with a semi-supervised algorithm.
In this work, we test two different thresholding methods using the Mahalanobis distances as scoring technique. MixMatch is used as the \gls{SSDL} algorithm to train the model. In order to assess the filter the accuracy of the model using the filtered unlabeled dataset, the filter is tested under some controlled setting to evaluate the accuracy obtained after training the model using the data selected by the filter. The proposed method is tested using real-world data with distribution mismatch conditions in a medical imaging analysis such as the COVID-19 detection using chest X-ray images.

\begin{table*}
\caption{COVID-19$^{-}$ observation sources description used in this work.}
\center
\resizebox{14cm}{!}{
\begin{tabular}{c|c|c|c|c|c}
\textbf{}                   & \textbf{CR-Chavarria-ChestXray-2020}                           & \textbf{Chinese dataset}                       & \textbf{ChestX-ray8 dataset}              & \textbf{Indiana dataset}                            \\ \hline
No. of patients             & 105                                            & 5856                                           & 65240                                     & 4000                                                                        \\
Patient's age range (years) & 7 -- 86                                           & children                                       & 0 -- 94                                      & adults                                                                     \\
No. of  obs.                & 105                                            & 5236                                           & 224316                                    & 8121                                                                        \\
Hospital/clinic             & Clinica Chavarria                              & No info.                                       & Stanford Hospital                         & Indiana Network                             &                    \\
                            &                                                &                                                &                                           & for Patient Care                                                        \\
Im. resolution              & $1907\times1791$                               & $1300\times600$                                & $1024\times1024$                          & $1400\times1400$                                              \\
Reference                   & \cite{calderon2021correcting} & \cite{kermany2018identifying} & \cite{irvin2019chexpert} & \cite{demner2016preparing} 
\end{tabular}}
\label{tab:datset_description}
\end{table*}

\section{Dataset Description}

\begin{figure}
    \centering
    \includegraphics[scale=0.18]{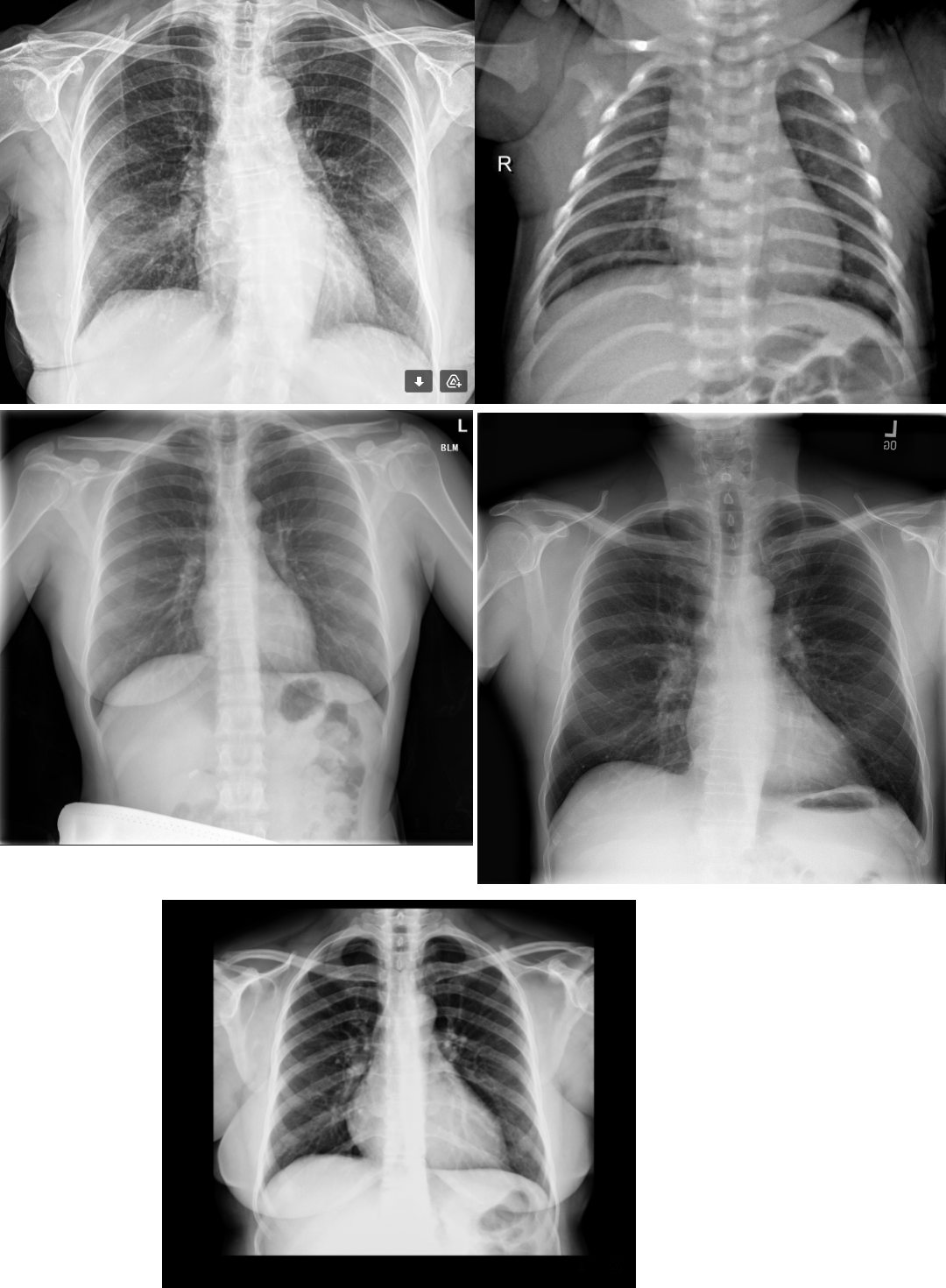}
    \caption{Row 1, column 1: a COVID-19$^{+}$ observation from \cite{cohen2020covid}, row 1, column 2:  a  COVID-19$^{-}$ observation from the Chinese dataset \cite{kermany2018identifying}, row 2, column 1: ChestX-ray8  COVID-19$^{-}$ image \cite{irvin2019chexpert}, row 2, column 2: Indiana dataset COVID-19$^{-}$  sample image \cite{demner2016preparing}. The bottom image corresponds to a sample image from the Costa Rica dataset \cite{calderon2021correcting}.  As it can be seen, images from the Costa Rica dataset originally include a black frame.}
    \label{fig:sample_images_chest}
\end{figure}

In this work, we asses the negative effect of distribution mismatch between the labeled and unlabeled dataset using the MixMatch algorithm.  We use the medical imaging domain to test the proposed methods under real-world usage conditions, such as  COVID-19 detection using chest X-ray images (binary classification between no pathology and pathology). We use different sources of COVID-19$^-$ (no pathology) and COVID-19$^+$(positive pathology) to recreate different distribution mismatch conditions. The observations of COVID-19$^+$ were collected from the open dataset available in \cite{cohen2020covidProspective}. This dataset currently has 105 COVID-19$^+$ images. Images in the dataset do not have the same resolutions, which range from $400 \times 400$ to $2500 \times 2500$. Observations of COVID-19$^-$ were collected from four different data sources. Such data sources are summarized in Table \ref{tab:datset_description}.  Figure \ref{fig:sample_images_chest} shows a sample of each dataset.

 For the COVID-19$^+$ images \cite{cohen2020covidProspective}, the observations related to Middle East Syndrome (MERS), Acute Respiratory Distress Syndrome (ARDS), and Severe Acute Respiratory Syndrome (SARS) were removed and the letters in the radiographs were cropped. In the case of the Indiana dataset the lateral projections were discarded. The Indiana and Chinese datasets contained observations of different pathologies, therefore, we preserved only observations categorized as normal.

The configurations for the experiments were done using different combination of labeled and unlabeled datasets, this is shown in Table \ref{tab:results1}. The experiments were carried out using $n_l=20$ labeled samples and $n_u=90$ for the unfiltered unlabeled datasets. We used different sources for the labeled dataset and different contamination level to asses the impact of distribution mismatch in an unlabeled dataset. As the different sources of data were datasets of COVID-19$^-$, the unlabeled datasets contained only COVID-19$^-$ observations. 

Different combinations of data sources were used to build the unlabeled dataset to create different distribution mismatch conditions. As we noticed in our empirical tests that the Costa Rican dataset yielded the lowest performance when used for \gls{SSDL} training, we used it as a \gls{OOD} contamination source. The percentage of contamination in the test-bed range from 20 to 100 percent, as seen in Table \ref{tab:results1}.  

Regarding the test dataset, it consists of observations from the same source as the labeled dataset. Hence, it contains observations of COVID-19$^+$ and its respective COVID-19$^-$ source. The test dataset used a total of 40 observations, with 20 observations per class.

% \todo{ELIMINAR VALENCIA}
\begin{table*}
\centering
\caption{Results: Accuracy of an Alexnet model trained with MixMatch, with different unlabeled datasets. The unlabeled datasets Costa Rican, ChestX-ray8, and Chinese dataset include only COVID-19$^-$ observations.}
\resizebox*{!}{.37\textwidth}{ %
    \begin{tabular}{l|l|l}
    \toprule
    \multicolumn{1}{c|}{\textbf{Labeled dataset $n_l=20$}} & \multicolumn{1}{l|}{\textbf{Unlabeled dataset $n_u=90$}} & \multicolumn{1}{c}{\textbf{Accuracy}} \\ \hline
    \multicolumn{1}{c|}{\multirow{7}{*}{Indiana}} & Supervised & $67.2  \pm  1.7$\\ %\cline{2-3}
    \multicolumn{1}{l|}{} & Indiana                                    & 
    $68.7  \pm  2.9$                       \\
    \multicolumn{1}{l|}{} & Indiana $80\%$ - Costa Rica $20\%$             & $67.3  \pm  1.7$                       \\
    \multicolumn{1}{l|}{}                         & Indiana $60\%$ - Costa Rica $40\%$                          & $67.0  \pm  1.9$                     \\
    \multicolumn{1}{l|}{}                         & Indiana $40\%$ - Costa Rica $60\%$                         & $63.7  \pm  3.6$                      \\
    \multicolumn{1}{l|}{}                         & Indiana $20\%$ - Costa Rica $80\%$                         & $60.5  \pm  2.4$                      \\
    \multicolumn{1}{l|}{}                         & Costa Rica                                        & $58.5  \pm  1.7$   \\  \hline
    
    \multicolumn{1}{c|}{\multirow{6}{*}{Indiana}} & ChestX-ray8                                    & $72.8  \pm  2.6$                       \\
    \multicolumn{1}{l|}{} & ChestX-ray8 $80\%$ - Costa Rica $20\%$             & $65.8  \pm  2.3$                      \\
    \multicolumn{1}{l|}{}                         & ChestX-ray8 $60\%$ - Costa Rica $40\%$                          & $62.0  \pm  2.7$                     \\
    \multicolumn{1}{l|}{}                         & ChestX-ray8 $40\%$ - Costa Rica $60\%$                         & $60.0  \pm  1.9$                      \\
    \multicolumn{1}{l|}{}                         & ChestX-ray8 $20\%$ - Costa Rica $80\%$                         & $60.3  \pm  2.8$                      \\
    \multicolumn{1}{l|}{}                         & Costa Rica                                        & $58.5  \pm  1.2$   \\  \hline
    
    % \multicolumn{1}{c|}{\multirow{7}{*}{ChestX-ray8}} 
    % & Supervised & xxxxx  \pm  xxxxx\\ 
    % \multicolumn{1}{l|}{} & ChestX-ray8                                    & 0.533  \pm  0.025                       \\
    % \multicolumn{1}{l|}{} & ChestX-ray8 80\% - Costa Rica 20\%             & 0.523  \pm  0.013                     \\
    % \multicolumn{1}{l|}{}                         & ChestX-ray8 60\% - Costa Rica 40\%                          & 0.518  \pm  0.016                  \\
    % \multicolumn{1}{l|}{}                         & ChestX-ray8 40\% - Costa Rica 60\%                         & 0.515  \pm  0.032                    \\
    % \multicolumn{1}{l|}{}                         & ChestX-ray8 20\% - Costa Rica 80\%                         & 0.5  \pm  0.039                     \\
    % \multicolumn{1}{l|}{}                         & Costa Rica                                        & 0.5  \pm  0.032  \\ \hline

    \multicolumn{1}{c|}{\multirow{7}{*}{China}} 
    & Supervised & $65.5  \pm  9.0$ \\ %\cline{2-3}
    \multicolumn{1}{l|}{} & China                                    & $98.0  \pm  1.0$                       \\
    \multicolumn{1}{l|}{} & China $80\%$ - Costa Rica $20\%$             &  $96.5  \pm  2.3$                    \\
    \multicolumn{1}{l|}{}                         & China $60\%$ - Costa Rica $40\%$                          &  $95.5  \pm  2.4$                 \\
    \multicolumn{1}{l|}{}                         & China $40\%$ - Costa Rica $60\%$                         &   $94.8  \pm  1.3$                   \\
    \multicolumn{1}{l|}{}                         & China $20\%$ - Costa Rica $80\%$                         &  $96.7  \pm  2.0$                     \\
    \multicolumn{1}{l|}{}                         & Costa Rica &  $96.5  \pm  1.2$  \\ \bottomrule
    \end{tabular} %
}
\label{tab:results1}
\end{table*}
\begin{table*}
    \centering
    \caption{Results: Accuracy of an Alexnet model trained with MixMatch, with the filtered datasets using the two filtering proposed method: Otsu's thresholding method and K-Means clustering, and also the filter using the real threshold. Bold values represent the highest accuracy achieve by the filter using the Otsu's thresholding method or K-means in each setting.}
    \resizebox*{!}{.29\textwidth}{ %
    \begin{tabular}{l|l|l|l|l}
    \toprule
    \multicolumn{1}{c|}{\textbf{Labeled dataset $n_l=20$}} & \multicolumn{1}{l|}{\textbf{Unlabeled Dataset $n_u=90$}} & \multicolumn{1}{c|}{\textbf{Acc. K-Means}} & \multicolumn{1}{c|}{\textbf{Acc. Otsu}} & \multicolumn{1}{c}{\textbf{Acc. Real threshold}} \\ \hline
    % ------------
    \multicolumn{1}{c|}{\multirow{5}{*}{Indiana}} & Indiana                                    & $\bm{69.2  \pm  2.5}$             & $69.0  \pm  4.8$   &  $71.3  \pm  6.5$        \\
    % ------------
    \multicolumn{1}{l|}{}                         & Indiana $80\%$- Costa Rica $20\%$                          &  $\bm{64.0  \pm  4.6}$             &  $62.7  \pm  3.4$   & $68.0  \pm  5.3$       \\
    % ------------
    \multicolumn{1}{l|}{}                         & Indiana $60\%$- Costa Rica $40\%$                          &  $65.2  \pm  3.9$            &  $\bm{65.7  \pm  3.2}$    & $65.5  \pm  3.5$      \\
    % ------------
    \multicolumn{1}{l|}{}                         & Indiana $40\%$- Costa Rica $60\%$                        & $62.5  \pm  2.5$             &  $\bm{63.0  \pm  2.2}$    &  $65.5  \pm  3.8$      \\
    % ------------
    \multicolumn{1}{l|}{}                         & Indiana $20\%$- Costa Rica $80\%$                        &  $\bm{59.8  \pm  3.1}$            & $58.5  \pm  2.3$   & $58.3  \pm  5.8$        
    % ------------
    \\ \hline
    \multicolumn{1}{c|}{\multirow{5}{*}{Indiana}}
    % ----------------------------------------------------
    & ChestX-ray8                                    & $72.2  \pm  3.2$             & $\bm{73.3  \pm  3.7}$   & $73.7  \pm  2.8$       \\
    % ------------
    \multicolumn{1}{l|}{}                         & ChestX-ray8 $80\%$- Costa Rica $20\%$                          &  $68.3  \pm  2.5$             &   $\bm{69.3  \pm  2.2}$     & $69.0  \pm  2.5$     \\
    % ------------
    \multicolumn{1}{l|}{}                         & ChestX-ray8 $60\%$- Costa Rica $40\%$                          &  $\bm{68.8  \pm  2.0}$            &  $68.5  \pm  3.0$   & $69.2  \pm  2.5$       \\
    % ------------
    \multicolumn{1}{l|}{}                         & ChestX-ray8 $40\%$ - Costa Rica $60\%$                         & $60.2  \pm  3.1$             & $\bm{61.3  \pm  1.7}$   & $66.2  \pm  3.2$       \\
    % ------------
    \multicolumn{1}{l|}{}                         & ChestX-ray8 $20\%$ - Costa Rica $80\%$                         & $\bm{60.0  \pm  2.7}$         & $58.5  \pm  2.8$ & $61.2  \pm  6.0$ \\ \hline        
    % ----------------------------------------------------
    
    % \multicolumn{1}{c|}{\multirow{5}{*}{ChestX-ray8}} 
    % & ChestX-ray8                                    & 0.528  \pm  0.024             &  0.525  \pm  0.011         \\
    % % ------------
    % \multicolumn{1}{l|}{}                         & ChestX-ray8 80\%- Costa Rica $20\%$                          &  0.52  \pm  0.019             &  0.53  \pm  0.015          \\
    % % ------------
    % \multicolumn{1}{l|}{}                         & ChestX-ray8 60\%- Costa Rica $40\%$                          &  $0.517  \pm  0.016$            &  $0.52  \pm  0.01$          \\
    % % ------------
    % \multicolumn{1}{l|}{}                         & ChestX-ray8 $40\%$ - Costa Rica $60\%$                         & $0.495  \pm  0.056$             &   $0.5  \pm  0.045$          \\
    % % ------------
    % \multicolumn{1}{l|}{}                         & ChestX-ray8 $20\%$ - Costa Rica $80\%$                         &   $0.497  \pm  0.039$        &   $0.508  \pm 0.039$ \\ \hline
    
    % ----------------------------------------------------
    
    \multicolumn{1}{c|}{\multirow{5}{*}{China}} 
    & China                                    &  $98.2  \pm  1.6$             &  $\bm{98.7  \pm  1.3}$ & $98.0  \pm  1.0$        \\
    % ------------
    \multicolumn{1}{l|}{}                         & China $80\%$- Costa Rica $20\%$                          & $ 98.0  \pm  1.5  $           &  $\bm{99.3  \pm  1.1}$  & $98.2  \pm  1.1$       \\
    % ------------
    \multicolumn{1}{l|}{}                         & China $60\%$- Costa Rica $40\%$                          &  $\bm{99.5  \pm  1.0}$            &  $98.8  \pm  1.3$  & $98.7  \pm  1.3$        \\
    % ------------
    \multicolumn{1}{l|}{}                         & China $40\%$- Costa Rica $60\%$                        & $93.8  \pm  1.7$             &   $\bm{94.5  \pm  1.0}$    & $93.8  \pm  4.1$      \\
    % ------------
    \multicolumn{1}{l|}{}                         & China $20\%$- Costa Rica $80\%$                        &   $95.7  \pm  1.1$         &   $\bm{96.2  \pm  2.1}$ & $52.5  \pm  0.0$ \\
    \bottomrule
    \end{tabular}  %
    }
    \label{tab:results2}
\end{table*}

\section{Experimental Design}

\subsection{Evaluation of using distribution mismatch datasets on MixMatch's accuracy}
\label{testbed1}
The first experimentation test-bed is designed to assess the effect of using different unlabeled dataset and a target labeled datasets in the MixMatch's performance. Hence, the prepared settings follow a controlled environment where the unlabeled datasets depicts different degrees of distribution mismatch.  Table \ref{tab:results1} shows the results of training MixMatch using an AlexNet model.

\subsection{Evaluation of MixMatch's accuracy using the Otsu's and K-means methods for automatic thresholding}
\label{testbed2}
The experiment is designed to asses the MixMatch's accuracy results when using the proposed methods to filter the \gls{OOD} data in the unlabeled dataset. Hence, the MixMatch's accuracy is measured using the filtered datasets with an AlexNet model. Table \ref{tab:results2} shows the results of training the AlexNet model using MixMatch with the filtered datasets by the proposed methods. The filtering methods use the Mahalanobis distance to assign a score to each unlabeled sample. As previously mentioned, we assume that the \gls{OOD} and \gls{IOD} data tend to form a Gaussian mixture distribution or two clusters. Therefore the \gls{OOD} data is filtered using the threshold estimated by Otsu's thresholding method and the K-means clustering. Moreover, the proposed method evaluates whether the unlabeled dataset depicts contamination using the coefficient of variation of each cluster. To evaluate the ideal setting where all the \gls{OOD} data is removed from the unlabeled dataset, we evaluate the accuracy yielded by using the real threshold (percentage of contamination using the Costa Rican dataset). The MixMatch algorithm is used with the recommended parameters in \cite{berthelot2019mixmatch}. The models were trained for 50 epochs with 10 random data partitions, each with each respective training and test dataset.

\section{Results}
The results from the experiment in section \ref{testbed1} are presented in Table \ref{tab:results1}. We can see how the different settings of unlabeled data affects the accuracy yielded by the MixMatch algorithm with an Alexnet back-bone.

For instance, the supervised model using a labeled dataset from Indiana and China yielded an accuracy of $67.2$ and $65.5$ respectively. The accuracy of these models was improved by using unlabeled data to $68.7$ and $98.0$ respectively. The unlabeled Costa Rica dataset yields the lowest accuracy when used along the labeled Indiana dataset. From there, in general,  the accuracy is affected as we injected observations in the unlabeled dataset from the Costa Rican data source. This accuracy degradation can be attributed to a different distribution between the labeled and unlabeled datasets.

Regarding the two proposed methods for automatic thresholding of the \gls{OOD} data, described in Table \ref{tab:results2}, we can see how, in general, both automatic thresholding methods tend to yield similar results when using the real unlabeled data contamination percentage. This suggests that both thresholding methods reach a similar \gls{OOD} score to discard the unlabeled data, than using the real contamination value. The Otsu method, which assumes a Gaussian mixture distribution of the data, yields slightly better accuracy gains when used to filter the unlabeled data, however, with no statistical significance.  Both methods can be considered a good alternative to automatically calculate the threshold to discard \gls{OOD} data.

\section{Conclusions}
In this work, we analyzed the impact of training a \gls{SSDL} model using a labeled and unlabeled dataset with distribution mismatch. The experiments were assessed under the medical imaging domain for COVID-19 detection. As an extension to the work in \cite{calderon2022dealing}, we propose two simple methods to estimate the best threshold to discard unlabeled data, using  \gls{OOD} score  proposed in \cite{calderon2022dealing}. 

The two thresholding methods evaluated in this work performed similarly, close to  the accuracy yielded by filtering the unlabeled data with the real contamination percentage. Therefore, any of the tested methods can be used to find the best \gls{OOD} score threshold.

The presented approach is meant to improve data quality before training the model. The approach is built upon a pre-trained model with Imagenet using the Mahalanobis distance in the feature space, as this configuration showed better results in \cite{calderon2022dealing}. The implemented proposed methods can be used to automatically filter harmful observation from the dataset when the unlabeled dataset is not highly contaminated. Furthermore, we proved that \gls{SSDL} may improve the results of \gls{SDL} model when the labeled dataset is small. 

The approach does not require to train a deep learning model, which makes it less expensive  alternative to increase the data quality in the unlabeled dataset.

\bibliographystyle{plain}
\bibliography{bibliography}

\begin{thebibliography}{10}

\bibitem{Arora2014}
Richa Arora.
\newblock {The training and practice of radiology in India: current trends.}
\newblock {\em Quantitative Imaging in Medicine and Surgery}, 4(6):449--44950,
  2014.

\bibitem{bachman2014learning}
Philip Bachman, Ouais Alsharif, and Doina Precup.
\newblock Learning with pseudo-ensembles.
\newblock In {\em Advances in Neural Information Processing Systems}, pages
  3365--3373, 2014.

\bibitem{berthelot2019remixmatch}
David Berthelot, Nicholas Carlini, Ekin~D Cubuk, Alex Kurakin, Kihyuk Sohn, Han
  Zhang, and Colin Raffel.
\newblock Remixmatch: Semi-supervised learning with distribution alignment and
  augmentation anchoring.
\newblock {\em arXiv preprint arXiv:1911.09785}, 2019.

\bibitem{berthelot2019mixmatch}
David Berthelot, Nicholas Carlini, Ian Goodfellow, Nicolas Papernot, Avital
  Oliver, and Colin~A Raffel.
\newblock Mixmatch: A holistic approach to semi-supervised learning.
\newblock {\em Advances in Neural Information Processing Systems}, 32, 2019.

\bibitem{calderon2021improving}
Saul Calderon-Ramirez, Diego Murillo-Hernandez, Kevin Rojas-Salazar,
  Luis-Alexander Calvo-Valverde, Shengxiang Yang, Armaghan Moemeni, David
  Elizondo, Ezequiel Lopez-Rubio, and Miguel Molina-Cabello.
\newblock Improving uncertainty estimations for mammogram classification using
  semi-supervised learning.
\newblock In {\em Institute of Electrical and Electronics Engineers}, 2021.

\bibitem{calderon2022_ssdl_mismatch}
Saul Calderon-Ramirez, Shengxiang Yang, and David Elizondo.
\newblock Semi-supervised deep learning for image classification with
  distribution mismatch: A survey.
\newblock {\em IEEE Transactions on Artificial Intelligence}, pages 1--15,
  2022.

\bibitem{calderon2022dealing}
Saul Calderon-Ramirez, Shengxiang Yang, David Elizondo, and Armaghan Moemeni.
\newblock Dealing with distribution mismatch in semi-supervised deep learning
  for covid-19 detection using chest x-ray images: A novel approach using
  feature densities.
\newblock {\em Applied Soft Computing}, 123:108983, 2022.

\bibitem{calderon2021correcting}
Saul Calderon-Ramirez, Shengxiang Yang, Armaghan Moemeni, David Elizondo, Simon
  Colreavy-Donnelly, Luis~Fernando Chavarr{\'\i}a-Estrada, and Miguel~A
  Molina-Cabello.
\newblock Correcting data imbalance for semi-supervised covid-19 detection
  using x-ray chest images.
\newblock {\em Applied Soft Computing}, 111:107692, 2021.

\bibitem{cohen2020covid}
Joseph~Paul Cohen, Paul Morrison, and Lan Dao.
\newblock Covid-19 image data collection.
\newblock {\em arXiv 2003.11597}, 2020.
\newblock Available at
  \url{https://github.com/ieee8023/covid-chestxray-dataset}.

\bibitem{cohen2020covidProspective}
Joseph~Paul Cohen, Paul Morrison, Lan Dao, Karsten Roth, Tim~Q Duong, and
  Marzyeh Ghassemi.
\newblock Covid-19 image data collection: Prospective predictions are the
  future.
\newblock {\em arXiv 2006.11988}, 2020.

\bibitem{demner2016preparing}
Dina Demner-Fushman, Marc~D Kohli, Marc~B Rosenman, Sonya~E Shooshan, Laritza
  Rodriguez, Sameer Antani, George~R Thoma, and Clement~J McDonald.
\newblock Preparing a collection of radiology examinations for distribution and
  retrieval.
\newblock {\em Journal of the American Medical Informatics Association},
  23(2):304--310, 2016.

\bibitem{doersch2015unsupervised}
Carl Doersch, Abhinav Gupta, and Alexei~A Efros.
\newblock Unsupervised visual representation learning by context prediction.
\newblock In {\em Proceedings of the IEEE International Conference on Computer
  Vision}, pages 1422--1430, 2015.

\bibitem{dong2018tri}
WeiWang Dong-DongChen and Zhi-HuaZhou WeiGao.
\newblock Tri-net for semi-supervised deep learning.
\newblock In {\em Proceedings of twenty-seventh international joint conference
  on artificial intelligence}, pages 2014--2020, 2018.

\bibitem{gal2016dropout}
Yarin Gal and Zoubin Ghahramani.
\newblock Dropout as a bayesian approximation: Representing model uncertainty
  in deep learning.
\newblock In {\em international conference on machine learning}, pages
  1050--1059. PMLR, 2016.

\bibitem{gonzalez2008digital}
Rafael~C. Gonzalez and Richard~E. Woods.
\newblock {\em Digital image processing}.
\newblock Prentice Hall, Upper Saddle River, N.J., 2008.

\bibitem{hendrycks2016baseline}
Dan Hendrycks and Kevin Gimpel.
\newblock A baseline for detecting misclassified and out-of-distribution
  examples in neural networks.
\newblock {\em arXiv preprint arXiv:1610.02136}, 2016.

\bibitem{hu2021simple}
Zijian Hu, Zhengyu Yang, Xuefeng Hu, and Ram Nevatia.
\newblock Simple: similar pseudo label exploitation for semi-supervised
  classification.
\newblock In {\em Proceedings of the IEEE/CVF Conference on Computer Vision and
  Pattern Recognition}, pages 15099--15108, 2021.

\bibitem{irvin2019chexpert}
Jeremy Irvin, Pranav Rajpurkar, Michael Ko, Yifan Yu, Silviana Ciurea-Ilcus,
  Chris Chute, Henrik Marklund, Behzad Haghgoo, Robyn Ball, Katie Shpanskaya,
  et~al.
\newblock Chexpert: A large chest radiograph dataset with uncertainty labels
  and expert comparison.
\newblock In {\em Proceedings of the AAAI Conference on Artificial
  Intelligence}, volume~33, pages 590--597, 2019.

\bibitem{kermany2018identifying}
Daniel~S Kermany, Michael Goldbaum, Wenjia Cai, Carolina~CS Valentim, Huiying
  Liang, Sally~L Baxter, Alex McKeown, Ge~Yang, Xiaokang Wu, Fangbing Yan,
  et~al.
\newblock Identifying medical diagnoses and treatable diseases by image-based
  deep learning.
\newblock {\em Cell}, 172(5):1122--1131, 2018.

\bibitem{laine2016temporal}
Samuli Laine and Timo Aila.
\newblock Temporal ensembling for semi-supervised learning.
\newblock {\em arXiv preprint arXiv:1610.02242}, 2016.

\bibitem{lecun2015deep}
Yann LeCun, Yoshua Bengio, and Geoffrey Hinton.
\newblock Deep learning.
\newblock {\em nature}, 521(7553):436--444, 2015.

\bibitem{lee2018simple}
Kimin Lee, Kibok Lee, Honglak Lee, and Jinwoo Shin.
\newblock A simple unified framework for detecting out-of-distribution samples
  and adversarial attacks.
\newblock In {\em Advances in Neural Information Processing Systems}, pages
  7167--7177, 2018.

\bibitem{oliver2018realistic}
Avital Oliver, Augustus Odena, Colin~A Raffel, Ekin~Dogus Cubuk, and Ian
  Goodfellow.
\newblock Realistic evaluation of deep semi-supervised learning algorithms.
\newblock {\em Advances in neural information processing systems}, 31, 2018.

\bibitem{Otsu1979}
N.~Otsu.
\newblock A threshold selection method from gray-level histograms.
\newblock {\em IEEE Transactions on Systems, Man and Cybernetics}, 9(1):62--66,
  January 1979.

\bibitem{ouali2020overview}
Yassine Ouali, C{\'e}line Hudelot, and Myriam Tami.
\newblock An overview of deep semi-supervised learning.
\newblock {\em arXiv preprint arXiv:2006.05278}, 2020.

\bibitem{simeoni2021rethinking}
Oriane Sim{\'e}oni, Mateusz Budnik, Yannis Avrithis, and Guillaume Gravier.
\newblock Rethinking deep active learning: Using unlabeled data at model
  training.
\newblock In {\em 2020 25th International Conference on Pattern Recognition
  (ICPR)}, pages 1220--1227. IEEE, 2021.

\bibitem{sohn2020fixmatch}
Kihyuk Sohn, David Berthelot, Nicholas Carlini, Zizhao Zhang, Han Zhang,
  Colin~A Raffel, Ekin~Dogus Cubuk, Alexey Kurakin, and Chun-Liang Li.
\newblock Fixmatch: Simplifying semi-supervised learning with consistency and
  confidence.
\newblock {\em Advances in Neural Information Processing Systems}, 33, 2020.

\bibitem{tarvainen2017mean}
Antti Tarvainen and Harri Valpola.
\newblock Mean teachers are better role models: Weight-averaged consistency
  targets improve semi-supervised deep learning results.
\newblock In {\em Advances in neural information processing systems}, pages
  1195--1204, 2017.

\bibitem{van2020simple}
Joost van Amersfoort, Lewis Smith, Yee~Whye Teh, and Yarin Gal.
\newblock Simple and scalable epistemic uncertainty estimation using a single
  deep deterministic neural network.
\newblock {\em CoRR}, abs/2003.02037, 2020.

\bibitem{van2020survey}
Jesper~E Van~Engelen and Holger~H Hoos.
\newblock A survey on semi-supervised learning.
\newblock {\em Machine Learning}, 109(2):373--440, 2020.

\bibitem{yu2020multi}
Qing Yu, Daiki Ikami, Go~Irie, and Kiyoharu Aizawa.
\newblock Multi-task curriculum framework for open-set semi-supervised
  learning.
\newblock In {\em European Conference on Computer Vision}, pages 438--454.
  Springer, 2020.

\bibitem{zhu2005semi}
Xiaojin~Jerry Zhu.
\newblock Semi-supervised learning literature survey.
\newblock Technical report, University of Wisconsin-Madison Department of
  Computer Sciences, 2005.

\end{thebibliography}

% that's all folks
\end{document}